# Polymeric Assembly of Gluten Proteins in an Aqueous Ethanol Solvent

Mohsen Dahesh[1,2,3], Amélie Banc[1,2], Agnès Duri[3], Marie-Hélène Morel[3], and Laurence Ramos[1,2]*

[1] Université Montpellier 2, Laboratoire Charles Coulomb UMR 5221, F-34095, Montpellier, France
[2] CNRS, Laboratoire Charles Coulomb UMR 5221, F-34095, Montpellier, France
[3] UMR IATE, UM2-CIRAD-INRA-SupAgro, 2 pl Pierre Viala, 34070 Montpellier, France.

* Laurence.ramos@univ-montp2.fr

## Abstract

The supramolecular organization of wheat gluten proteins is largely unknown due to the intrinsic complexity of this family of proteins and their insolubility in water. We fractionate gluten in a water/ethanol (50/50 v/v) and obtain a protein extract which is depleted in gliadin, the monomeric part of wheat gluten proteins, and enriched in glutenin, the polymeric part of wheat gluten proteins. We investigate the structure of the proteins in the solvent used for extraction over a wide range of concentration, by combining X-ray scattering and multi-angle static and dynamic light scattering. Our data show that, in the ethanol/water mixture, the proteins display features characteristic of flexible polymer chains in a good solvent. In the dilute regime, the protein form very loose structures of characteristic size 150 nm, with an internal dynamics which is quantitatively similar to that of branched polymer coils. In more concentrated regimes, data highlight a hierarchical structure with one characteristic length scale of the order of a few nm, which displays the scaling with concentration expected for a semi-dilute polymer in good solvent, and a fractal arrangement at much larger length scale. This structure is strikingly similar to that of polymeric gels, thus providing some factual knowledge to rationalize the viscoelastic properties of wheat gluten proteins and their assemblies.

# 1. Introduction

Wheat storage gluten proteins are among the most complex proteins families comprising at least fifty different proteins, with extremely broad polymorphisms [1]. Wheat gluten proteins are moreover largely insoluble in water, rendering their study difficult. They are subdivided into two groups on the basis of their ability to form polymeric species by means of intermolecular disulfide bonds between protein subunits [2]. Gliadins, which account for about half of the total gluten proteins, are monomeric proteins and have molecular weight $M_w$ in the range 30-80 kDa. The other half are polymeric glutenins, modeled as a linear backbone of glutenin sub-units (with $M_w$ ranging from 30 kDa to 90 kDa) linked together by disulfide bonds [3, 4]. Gluten proteins belong to the family of disordered elastomeric proteins [5] which comprise among others elastin from vertebrates, abductin from arthropods, and dragline and flagelliform silks from spiders. All these proteins share common features in their primary structure as they all possess unstructured repeated domains rich in glycine and proline [6]. As opposed to most elastomeric proteins, the repetitive domains of gluten proteins are also rich in glutamine, which is prone to hydrogen-bonding. Gluten proteins are the most documented elastomeric plant proteins [7], and are responsible for the remarkable viscoelastic properties of dough. It is traditionally admitted that gliadins provide viscosity to the dough whereas elasticity originates from intermolecular interactions, through disulfide bonds and/or hydrogen bonds, between glutenins polymers [8-11]. In its water hydrated state, gluten forms a cohesive viscoelastic mass which is crucial in food science as it allows wheat flour to be baked into bread and biscuit or processed into pasta and noodles, but can also be manufactured into biomaterials [12]. Despite extensive studies over more than 200 years, in order to provide structural and mechanistic basis for the improvement of the viscoelastic properties of dough and of the quality of resulting food products, there is still a need to understand supramolecular organization [13]. This is partly due to the fact that these proteins do not crystallize and have an ill-defined three-dimensional structure determined majorly by indirect methods.

Attempts have been made to derive structural information from the viscoelastic response of gluten. Two conflicting views on the structure of gluten gels are described in the literature [14]. One of those considers that dense aggregates of gluten proteins form a percolated network as would do attractive colloidal particles [15-18]. On the other hand, gluten gels could be regarded as polymeric gels with junction points between flexible chains [9, 19-21]. Careful rheology experiments have shown that hydrated gluten and wheat flour doughs exhibit the linear visco-elastic response of critical gels, strongly supporting the second view of a gluten gel being a polymeric network [22-24]. Most experiments have been done in experimental conditions closed to those followed for making dough (typically 0.6-0.8 g of water per gram of flour). When performed on gluten, experiments were done in a rather concentrated regime (1.7g of water per gram of gluten) [24]. However, gluten proteins are not water soluble rendering hazardous a structural deduction of the polymer network from a mechanical response as out-of-equilibrium processes dominate.

Our approach is by contrast to directly probe gluten proteins structure in a food-grade solvent (an equi-volume mixture of water and ethanol) in which those proteins can be dispersed but nevertheless display a spontaneous gelation, yielding gel similar to gluten, when they are sufficiently concentrated. Thanks to a proper dispersion, a wide range of protein concentration is investigated, from very dilute solutions to concentrated ones. We used a combination of several scattering techniques, including static and dynamic light scattering, X-ray and neutron scattering, to probe the structure of gluten protein in solution. We show that our data can be rationalized in the framework of classical theory of polymers in a good solvent. Our data suggest also that the polymers are branched and form at large concentration polymeric gels with large length scale heterogeneities.

The paper is organized as follows. We first describe the protocol used to fractionate the proteins and briefly expose the characterization of the protein extract, and its behavior in aqueous ethanol solutions. We then probe the structure and dynamic of the protein suspension in the dilute regime, and in a concentrated regime using a combination of several scattering techniques. We finally discuss our results and compare our findings with previous investigations on the assembly of gluten proteins.

## 2. Materials and Methods

### 2.1. Materials

Ethanol and all chemicals were of analytical grade. Native gluten powder (81.94% protein, dry basis) was courtesy of Tereos-Syral (France).

*Gluten protein fractionation*
Our aim was to isolate a soluble protein fraction representative of the native gluten protein composition. The protocol was derived from the wheat flour protein fractionation developed by Boire *et al.* [25] and was similar to the one described in [26]. Gluten powder (20g) was placed in a centrifuge bottle (volume 250 mL) with 200 mL of 50 % (v/v) ethanol/water gluten and submitted to a continuous rotating agitation (60 rpm at 20°C) for 19 hours. After a 30 minutes centrifugation at 15000g at 25°C, the clear supernatant (S1) was recovered and placed at 4°C for 24 hours. This led to a phase separation between a light (S2) and a dense phase (C2). The latter, once recovered, was immediately frozen at -18°C before being freeze-dried and grinded. Global protein yield in C2, the extract of interest, was typically about 17%.

*Protein characterization*
Protein composition of the different extracts was checked by SDS-PAGE using the classical Laemmli buffer system (Tris-HCL/Tris-glycine). The 1.5 mm width gel comprised 120 mm of resolving gel (T = 13%, C = 0.63%) and 40 mm of stacking gel (T = 5.6%, C = 10%). Protein extracts (30 mg/mL) were suspended in Tris-HCl buffer, 2% sodium dodecyl sulphate (SDS), 13% glycerol and reduced with β-mercaptoethanol (5%) during at least 2 hours prior to loading (3 μL). Protein molecular weight standards (PageRuler$^{TM}$ unstained protein lader, Thermo Fisher) were also applied according to the instruction of the manufacturer. The gel was run at 40mA at 18°C and stopped half an hour after the leading front had reached the bottom of the gel. The gel was stained during 48h in 13% (w/v) trichloroacetic acid, 0.04% (w/v) Coomassie Brilliant Blue R250, and destained for 3 days with 10% trichloroacetic acid.

Protein size distribution was checked using Size Exclusion High-Performance Liquid Chromatography (SE-HPLC) performed on an Alliance system controlled by the Millenium software (Waters). Known quantity of samples (mass of C2 powder or volume of S1 and S2 solutions) were gently mixed 1 hour at room temperature with ethanol/water (50/50, v/v) at approximately 1 mg/ml before being diluted once in the elution buffer enriched at 1% SDS. Gluten powder was dispersed at 0.8 mg/mL in SDS buffer and mixed 80 min at 20°C. Two columns placed in series (TSK gel G6000 and G5000 PWXL, each 30 cm x 7.8 mm I.D., TosoBioscience) and preceded by a guard-column (TSK gel PWXL, 4 cm x 6 mm I.D., TosoBiocience) were used. Elution of the injected sample (20μl) was performed at 0.5 ml/min with 0.1% sodium dodecyl sulfate, 0.1 M sodium phosphate buffer at pH 6.8 (elution buffer) with detection of the different species at a wavelength of 214 nm. The protein concentrations of the injected samples were estimated from the corresponding SE-HPLC total areas using previously determined calibration factor [27]. The size-exclusion system was calibrated using blue dextran (2000 kDa), thyroglobulin (670 kDa), bovine serum albumin (66.5 and 132 kDa), catalase (116 kDa and 58 kDa), carbonic anhydrase (29 kDa), trypsinogen (23.7 kDa), cytochrome C (13.3 kDa), and dithionitrobenzoic acid for total column volume estimation. The above protein samples (S1, S2, C2 and gluten) were also analyzed using the same elution buffer on a TSK G400 SWXL column (30 cm x 7.8 mm I.D., TosoBioscience). With this column, albumin/globulin (25 kDa-10 kDa) and gliadin (90-25 kDa) that were confused on the PWXL columns can accurately be separated and quantified [28].

*Sample preparation*
The samples were prepared by dispersing the required mass of proteins (C2) in the appropriate solvent, an equi-volume mixture of ethanol and milliQ water. The protein/solvent mixtures were put in a rotary shaker overnight at room temperature to allow sample homogenization. They were subsequently kept at 20°C and used within 15 days after preparation.

## 2.2. Experimental methods

*Turbidity*

The turbidity of the samples was determined using a UV-VIS-NIR Varian 5000 spectrometer at room temperature for wavelength in the range (400–800) nm where the proteins do not absorb light. The results are given as the absorbance, defined as $A = -\log(I_t/I_0)$, where $I_t$ (resp. $I_0$) are the transmitted intensities through the sample (resp. solvent), measured at 600 nm. The samples were dissolved in ethanol/water (v/v 50%) mixture and placed into 1 mm thick quartz cells. Before each measurement two empty cells were inserted into sample holders in order to get blank spectra. For measurements, one cell was filled with pure solvent ethanol/water (v/v 50%) and used as reference, and the other cell was filled with the sample. Measurements were performed at typically 5 positions along the height of the cell. Results are given as average over the different measurements with error bars corresponding to the standard deviation of the different measurements.

*Light scattering measurements*

Static and dynamic and light scattering measurements were carried out for scattering angle $\theta$ in the range (20-140) deg, yielding scattering wave vectors $q = 4\pi n/\lambda \sin(\theta/2)$ in the range (5.6-30.1)x$10^6$ m$^{-1}$. Here $n$=1.357 is the refractive index of the solvent, and $\lambda$= 532 nm is the wavelength of the incident light (a Cobalt diode-pumped solid-state laser). Before measurements samples were filtered using a hydrophilic (surfactant-free cellulose acetate) syringe filter with a pore size of 0.8μm to remove dust particles. We checked by SEC-HPLC that filtration did not have any effect on the molecular size distribution. The concentration after filtration was measured by UV-vis spectrometry using the peak of protein absorption at 277 nm, considering an extinction coefficient of 0.565 ml/mg/cm (as measured by UV-vis spectroscopy for solutions comprising protein concentrations in the range (0.3-4.0) mg/ml).

For static light scattering measurements, an analyzer with polarization parallel to that of the incident light was placed in front of the detector. The excess Rayleigh ratio, $\mathcal{R}_{ex}$, was measured thanks to standard procedures using toluene for absolute calibration [29]. In brief, $\mathcal{R}_{ex} = \mathcal{R}_{std} \frac{I_s - I_{sol}}{I_{std}}$, where $I_s$ is the intensity scattered by the sample, $I_{sol}$ is the intensity scattered by the solvent, and $I_{std}$ is the intensity scattered by toluene, and $\mathcal{R}_{std}$= 2.5373×10$^{-5}$ cm$^{-1}$ is the Rayleigh ratio of toluene at 532 nm [30]. Measurements were performed at different scattering angles, and for various sample concentrations, $C$. We use Berry analysis, where $\left(\frac{KC}{\mathcal{R}_{ex}}\right)^{1/2}$ is plotted as a function of $\sin^2(\theta/2) + C$. Here $C$ is the protein concentration, and $K = 4\pi^2 n^2 (dn/dC)^2 / \mathcal{N}_a \lambda^4$ is the scattering contrast with $\mathcal{N}_a$ the Avogadro number, and $dn/dC = 0.149$ ml/g the refractive index increment, which was determined by measuring the index of refraction of solutions comprising protein concentration in the range (20-200) mg/ml. Note that Berry analysis has been found to be more accurate than a Zimm plot analysis for objects with radius of gyration of the order or larger than 100 nm [31]. The apparent molecular weight of the scattering objects, $M_w^{app}$, the second virial coefficient, $A_2$, and radius of gyration can be evaluated from a Berry plot, through linear extrapolations of the data towards $C$=0 and $\theta$=0. A linear extrapolation of the data towards $C$=0 yields $(KC/\mathcal{R}_{ex})^{1/2} \to \frac{1}{(M_w^{app})^{1/2}}\left(1 + \frac{q^2 R_G^2}{6}\right)$, in the regime $qR_G \ll 1$. In this limit, the Berry plot shows a linear variation of $(KC/\mathcal{R}_{ex})^{1/2}$ vs $\sin^2(\theta/2) + C$, and the apparent molecular weight and the radius of gyration are directly evaluated from the intercept and slope of the linear function. On the other hand, a linear extrapolation towards $\theta$=0 yields $(KC/\mathcal{R}_{ex})^{1/2} \to \frac{1}{(M_w^{app})^{1/2}}\left(1 + A_2 M_w^{app} C\right)$. In this limit, the Berry plot shows a linear variation of $(KC/\mathcal{R}_{ex})^{1/2}$ vs $\sin^2(\theta/2) + C$, and the apparent molecular weight and the second virial coefficient are directly evaluated from the intercept and slope of the linear function.

The intensity auto-correlation functions, $g_2(\tau)$-1, were calculated using an Amtec goniometer and a Brookhaven BT9000 correlator for delay times, $\tau$, in the range (10$^{-7}$ - 3) s. They were analyzed using cumulant expansion $g_2(\tau) - 1 = \beta \times \exp(-2\Gamma_1\tau + K_2^2\tau^2)$, where $\beta \leq 1$ is an instrumental constant [32], $\Gamma_1$ is the first decay rate, and $K_2$ is the second cumulant moment. As a cross-check, the correlation functions were also analyzed by the inverse Laplace transform program CONTIN.

*X-ray scattering*

Synchrotron small-angle X-ray scattering (SAXS) experiments were conducted at the European Synchrotron radiation facilities (ESRF) at Grenoble, France on the ID2 beam line. The samples were held in capillaries of internal diameter 1.5 mm. The beam wavelength is 1 Å and the sample-to-detector distance is 2.5 m, yielding scattering wave-vectors in the range (0.04 - 1.5) nm$^{-1}$. The scattered intensity, *I(q)*, was obtained by using standard procedures, including background subtraction given by ESRF. Additional data were obtained using a laboratory set-up. In brief, a high brightness low power X-ray tube, coupled with aspheric multilayer optic (GeniX 3D from Xenocs) was employed, which delivers an ultralow divergent beam (0.5 mrad). Scatterless slits were used to give a clean 0.8 mm diameter X-ray spot with an estimated flux around 35 Mph/s at the sample position. The scattered intensity was collected on a two-dimensional Schneider 2D image plate detector prototype, at a distance of 0.2 m from the sample. The experimental data were corrected for the background scattering.

*Very small-angle neutron scattering*

Very small-angle angle neutron scattering experiments (VSANS) running on the focusing mirror principle [33] were performed at the KWS3 instrument operated by FRM II at the Heinz Maier-Leibnitz Zentrum (MLZ), Garching, Germany. The samples were held in 1-mm thick quartz cells. Two sample-to-detector distances (1.2 m and 9.5 m) were used with a wavelength of λ= 12.8 Å to access *q*-vectors from $2\ 10^{-3}$ to $10^{-1}$ nm$^{-1}$, allowing an overlap with the SAXS data. The reduction of raw data was performed by the routine qtiKWS [34] including corrections for detector sensitivity, background noise and empty cell signal. Note that, despite a low contrast between the proteins and the hydrogenated solvent, measurements were performed on samples prepared with the hydrogenated solvent as deuteration induced alterations of the scattering signal (data not shown).

## 3. Results and Discussion

### 3.1. Characterization of the gluten protein extract

Figure 1 presents the molecular weight-distribution profiles of native gluten, S1, S2 and C2 protein extracts, as obtained from Size Exclusion High-Performance Liquid Chromatography (SE-HPLC), while their protein subunits composition can be inferred from the reduced SDS-PAGE patterns given in Figure 2. In both cases, equivalent protein contents were used in order to exemplify the difference in protein composition. About 84.5% of the native gluten protein is found extractable in SDS-phosphate buffer at room temperature. This solvent is one of the most efficient to bring gluten into solution because SDS interacts with the polypeptide backbone disrupting ionic, hydrogen and hydrophobic bonds that stabilize the protein secondary through quaternary structure [35]. The protein residue (15.5% of total protein) can be assimilated to the largest glutenin polymers, which are well known to remain insoluble even in strong disrupting solvents [4]. SE-HPLC analysis shows that the molecular weight of the SDS-soluble glutenin polymers ranges from 6000 kDa to 90 kDa, while that of the later eluted gliadin ranges from 90 kDa to 11 kDa. The gliadin profile consists in a main peak flanked by a shoulder at higher Mw (smaller elution volume) (Fig. 1, blue dotted line). The gliadin profile can be better resolved using one column (Supplementary data, Fig. S1). From the quantitative analysis of SE-HPLC profile, native gluten is found to comprise 44.3% gliadin and 45.7% glutenin polymers, 30.2% of which being soluble in SDS-phosphate buffer. Extraction of gluten in ethanol/water (50/50, v/v) solubilized 48.5% of total gluten protein within S1 extract. As expected from the respective solubility of gliadin and glutenin in ethanol/water solvent, S1 is enriched in gliadin and depleted in glutenin polymers compare to native gluten (Fig. 1, purple dash line). From the SDS-PAGE patterns and compared to gluten, S1 is found to be enriched in α- and β-gliadins, and low-molecular-weight glutenin subunits (LMW-GS) while slightly depleted in high molecular weight glutenin subunits (HMW-GS) (Fig. 2). HMW-GS are known to be involved in the formation of the largest glutenin polymers [36], which are found relatively less abundant in S1 than in gluten (Fig. 1). The partitioning of S1 into dense (C2) and light (S2) phases by liquid-liquid phase separation at 4°C results in the disappearance of the largest glutenin polymers within the light phase (S2), while allowing their recovery within the dense phase (C2) (Fig. 1). SDS-PAGE analysis shows that the HMW-GS present in S1 are almost totally recovered within C2 while S2 is significantly enriched in α- and β-gliadins (Fig. 2). Careful examination of the size-distribution of glutenin polymers from C2 and gluten extracts shows that they are very similar (Supplementary data, Fig. S2). Finally, C2 showed a higher glutenin polymers/gliadin ratio than native gluten (1.11 against 1.01 or 0.66 depending on whether the native gluten SDS-insoluble glutenin polymers are taken into account or not in our estimate). Therefore, our fractionation method allows the isolation in 50% ethanol/water, the traditional solvent of gliadin, of a gluten proteins blend specifically enriched in glutenin polymers.

### 3.2. Protein solubility in ethanolic solvents

We have checked the solubility of the protein mixtures in different solvents, pure water, pure ethanol and ethanol water mixtures of different compositions (at different ratios). Homogeneous to the eyes mixtures are obtained for ratio of water over ethanol ranging from 40/60 to 60/40 (volume/volume), in agreement with experimental studies on wheat gluten proteins [37, 38]. In the following, we use a mixture of 50/50 of water/ethanol. Pictures of a samples series, comprising protein concentrations from $C$=10 mg/ml to $C$=600 mg/ml, are shown in Figure 3. All samples display a yellow color presumably originating from carotenoids. Note that samples with protein concentration larger than typically 150 mg/ml are viscoelastic gels.

As can be seen on Figure 3, the samples are more and more turbid as $C$ increases until reaching a plateau for $C$ in the range 245-320 mg/ml. For higher concentrations, the samples become more transparent. These observations are confirmed by turbidity measurements (Fig. 4). Note however that measurements were not possible for concentrations larger than 320 mg/ml, as we were not able to fill the cell without bubbles, due to the very high elasticity of the samples. Hence the decrease of the turbidity for high enough concentration hinted visually cannot be measured quantitatively.

### 3.3. Structure and dynamics in the dilute regime

Static light scattering measurements were performed in the dilute regime, for $C$ in the range (2-9) mg/ml. Figure 5a shows the variation of the scattering intensity, $R_{ex}/KC$, as a function of the wave-vector, $q$. The data shown for five concentrations almost overlap and display a plateau at low $q$ followed by a decrease of the scattered intensity at higher $q$. Data can be reasonably well fitted by a Debye function, which is the form factor for Gaussian chains [39, 40]: $f_D = \frac{2}{x^2}(e^{-x} - 1 + x)$, with $x = (qR_G)^2$. As can be seen on Figure 5a, this function, which is the exact form of the static form factor for a random walk polymer, accounts well for our experimental data in the whole range of accessible wave-vectors. From the fit of the data, the radius of gyration of the scattering objects is determined. We find within error bars a constant value for the five concentrations investigated: $R_G$=(147±6) nm. Note that a simpler Guinier analysis [39], where $\ln(R_{ex}/KC)$ vs $q^2$ is fitted with a linear function whose slope is $\frac{R_G^2}{3}$ could be used, in the regime where $qR_G \ll 1$. Smaller values, [$R_G$ = (117 ± 4) nm] are found in this case (data not shown). However the range of wave-vectors where the Guinier approximation holds is very restricted rendering the fit procedure not very reliable.

Additional information can be obtained on the structure of the samples by performing a Berry analysis, as detailed in the Materials and Methods section. We show in Figure 5b a Berry plot where $\left(\frac{KC}{\mathcal{R}_{ex}}\right)^{1/2}$ is plotted as a function of $\sin^2(\theta/2) + C$, for data acquired at different scattering angles, $\theta$, and for four different concentrations, from 2.1 mg/ml to 8.0 mg/ml. As explained in the Materials and Methods section, a linear extrapolation towards $\theta = 0$ and $C = 0$ allows one to evaluate the radius of gyration, second virial coefficient and apparent molecular weight of the scattering objects. From the Berry plot (Fig. 5b), one derives the following numerical values: radius of gyration $R_G = (149 \pm 3)$ nm, second virial coefficient $A_2 = 1.31 \times 10^{-6}$ ml mol/g$^2$, and apparent molecular weight of the scattering objects, $M_w^{app} = (26.83 \pm 0.16) \times 10^6$ g/mol. Note that this latter value is an average of the two values measured thanks the extrapolation towards $C=0$ and towards $\theta = 0$. Note that Berry analysis has been found to be more accurate than a Zimm plot analysis for objects with radius of gyration of the order or larger than 100 nm, as it is the case here [31]. We find here a radius of gyration comparable to the one measured by fitting the whole scattering curve (Fig. 5a), showing the consistency of our approach. We also measure a positive but very small value for $A_2$, as expected for big objects. For long polymer chains in a good solvent, the average volume occupied by one chain, $R_G^3$, is expected to be equal to $\frac{A_2 \times (M_w^{app})^2}{\mathcal{N}_a}$, with $\mathcal{N}_a$, the Avogadro number [41]. We estimate this way $R_G \approx 116$ nm, a numerical value in good agreement with the values measured above, suggesting that the protein assembly can be ascribed as polymer in good solvent conditions. This claim is confirmed by dynamic light scattering data.

To get more insight on the morphology of the scattering objects, dynamic light scattering experiments were performed. Intensity correlation functions were measured at several angles corresponding to several wave-vectors. Representative intensity correlation functions are shown in Figure 6. Data at low $q$ can be well fitted with a single exponential decay function, as shown in Figure 6, where fits of the data (lines in Fig. 6) acquired at $\theta = 35$ deg. ($q= 9.64\ 10^6$ m$^{-1}$) and $\theta = 50$ deg. ($q= 13.55\ 10^6$ m$^{-1}$) are shown. This proves that the scattering objects are close to monodisperse, which is rather unexpected in view of the broad size molecular weight distribution of the protein in the sample, as shown in the molecular weight profile (Fig. 2). The size distribution derived from a Contin analysis of the correlation functions at low $q$ also underlines the narrowness of the size distribution (inset Fig. 6). By contrast, a single exponential decay functional form fails to correctly account for the correlation function at large $q$, as shown in Figure 6 where the best fits with a single exponential decay are plotted together with the experimental data points for $\theta = 70$ deg. ($q = 18.40\ 10^6$ m$^{-1}$), $\theta = 90$ deg. ($q = 22.68\ 10^6$ m$^{-1}$), and $\theta = 130$ deg. ($q= 29.07\ 10^6$ m$^{-1}$). Because the shape of the correlation function changes with $q$, but remains always close to a mono-exponential, a cumulant analysis is chosen to analyze in a similar manner the correlation functions acquired in the whole range of wave-vectors. The first decay rate, $\Gamma_1$, is plotted as a function of $q^2$ in a lin-lin scale in Figure 7a and as a function of $q$ in a log-log scale in Figure 7b. Interestingly, we find that at low $q$, $\Gamma_1$ is proportional to $q^2$, as expected for a diffusive process. The proportionality factor is the translation diffusion constant $D$, which is related to the hydrodynamic radius of the objects via the Stokes-Einstein relation, $D = \frac{k_B T}{6\pi\eta_0 R_H}$, where $k_B T$ is the thermal energy and $\eta_0$ is the solvent viscosity ($\eta_0 = 2.788$ mPa s as measured with a capillary viscometer). We measure $R_H = (128 \pm 8)$ nm. Here the errors bars are the standard deviations of the results measured at different concentrations.

A faster than a $q^2$ dependence of the decay rate is however measured at higher $q$. This feature is the sign that, in addition to translation diffusion, other mechanisms occur, leading to a faster relaxation of the correlation functions. The departure from the $q^2$ dependence occurs roughly when $qR_G$ becomes larger than 1, hence when one probes the dynamic of the scattering objects at length scale smaller than their size. Our observations suggest therefore an internal dynamics of the scattering objects. Interestingly, our data present strong analogies with data obtained for linear or branched synthetic polymer coils [42-47], but also for more complex natural polymers [48]. At high $q$, the relaxation rate is found to vary as a power law with the wave vector, $\Gamma \sim q^m$. We find $m = 2.98 \pm 0.16$, where the error bars come from the average of the different concentrations (inset Fig. 7b). The power law is in agreement with the theoretical expectation and experimental observations for polymers. Indeed, an asymptotic power law with $m = 3$ has been predicted for chains with strong hydrodynamic interactions (Zimm limit) [49]. In this limit, the shape of the correlation functions has been calculated [42, 49]. At large $q$, such that $qR_G \gg 1$ and internal dynamics dominate, the asymptotic time-dependence of the correlation function is a stretched exponential decay $(g_2(q,\tau) - 1)/\beta \rightarrow \exp[-(\tau/t^*)^\alpha]$ with $\alpha=2/3$.

Note that $\alpha=1/2$ for Rouse relaxation without hydrodynamic interaction. At low $q$ by contrast, where the relaxation is dominated by the translational diffusion of the scattering objects, the correlation function, when the objects are monodisperse, is a simple exponential, as measured (Fig. 6). To emphasize the change of shape as the wave vector increases, the correlation functions are plotted as a function of a normalized delay, $\Gamma_1\tau$, where $\tau$ is the delay time and $\Gamma_1$ is the decay rate extracted from the cumulant analysis of the correlation functions (Fig. 8). By construction, all correlation functions taken at different angles collapse at small normalized delay but depart markedly at large $\Gamma_1\tau$. We find that the shape of the correlation functions departs notably from a simple exponential when $qR_G>2$. This cross-over corresponds to the $q$-vector below which translational diffusion is probed and above which internal mode dynamics is probed (Fig. 7). Consistently, we find for the largest angle that the shape of the correlation function reaches the one predicted in the Zimm limit.

In the Zimm limit, for polymers in good solvent, $\Gamma^* = \frac{\Gamma_1}{q^3}\frac{\eta_0}{k_BT}$ is expected to reach a constant $q$-independent value $\Gamma_\infty^* = 0.071 - 0.079$ when $qR_G$ (or $qR_H$) tends to infinity [43]. To quantitatively check this prediction, we have computed $\Gamma^*$ in the whole range of wave-vectors and for the five concentrations investigated. Figure 7c gathers our results and display $\Gamma^*$ as a function of $qR_H$. Interestingly, the data for the five concentrations investigated reasonably collapse onto a single curve, which exhibits two regimes. At low $q$ ($qR_H < 2$), the data follow the expected law for a purely translational motion ($\Gamma^* = \frac{1}{6\pi R_H q}$), whereas at higher $q$ ($qR_H > 2$), $\Gamma^*$ decays slower than the expected law for a purely translational motion and reaches a plateau, $\Gamma_\infty^* = 0.030 \pm 0.004$. This value is about two times lower than both the theoretical predictions and the experimental observations for linear chains in good solvent [43-47]. This indicates that some predicted internal motions are not observable. This could be due to a hyperbranched structure as invoked in [48] or to cross-links inside the objects, as it has been shown to greatly reduce $\Gamma_\infty^*$ compared to linear chains [50, 51]. Consistent conclusions are derived from the examination of the ratio of the radius of gyration over the hydrodynamic radius, $\rho = R_G/R_H$, as this ratio provides some information on the morphology of the scattering objects [52]. We find $\rho$ of the order of 1.15. This value is much higher than the one expected for a compact spherical object ($\rho = 0.775$) but smaller than the one obtained both theoretically and experimentally for a flexible chain in good solvent ($\rho$ in the range 1.3-1.7) [44, 45, 52]. Our result is therefore consistent with scattering objects being branched or cross-linked, as $\rho$ has been shown to decrease with branching [48, 51, 53].

### 3.4. Structure in the concentrated regime

X-ray and neutron scattering measurements have been performed for samples in a rather concentrated regime, with proteins concentration in the range (100-400) mg/ml. A combination of various instruments (VSANS, SAXS, WAXS) is used to probe the structure from atomic scale to mesoscopic scale. We demonstrate below that the whole scattering curve, scattered intensity *vs* scattering vectors, can be accounted by a model of polymeric gels in good solvent conditions. As an illustration, we show in Figure 9 the characteristic scattering for a sample with a protein concentration $C$=290 mg/ml. At high $q$, a wide peak centered at $q$=16.8 nm$^{-1}$ is observed. It corresponds to the liquid order of the sample displaying average distances between atoms of the order of 0.37 nm. The average position of this peak is a combination of protein and solvent liquid orders [54] and is not sensitive to the protein secondary structure [55]. Between 2 and 8 nm$^{-1}$, the scattering curve displays a $q^{-1}$ decrease. This power law variation is characteristic of rigid rods, and can model polymeric chains at a scale lower than the persistence length. In addition, at lower wave-vectors, a transition towards a stronger $q$-dependence is expected as one probes the flexible conformation of the polymer chains. We find that in the range (0.8-2) nm$^{-1}$, the scattering data can be well accounted for by a $q^{-5/3}$ decrease, which is the theoretical expectation for a self-avoiding polymer chain in a good solvent. The $q$-range over which the $q^{-5/3}$ scaling is measured is here narrow, because the sample is rather concentrated. However, this scaling extends over almost two orders of magnitude of wave-vectors for a more dilute sample, as shown in figure 10. The transition between the $q^{-1}$ regime and the $q^{-5/3}$ regime is predicted to occur at $q^*$ position related to the persistence length $l_0$ of the chain, such that $q^* \times l_0 = 1.9$ [56, 57]. Because the transition is very often rather broad, it is common to use a so-called "Holtzer plot" where $qI$ is plotted as a function of $q$. With this representation, the $q^{-1}$ dependence of an infinitely thin rigid rod results in a plateau (inset Fig. 10). We experimentally find $q^* \cong 2.6$ nm$^{-1}$, yielding a persistence length of the order of 0.7 nm. Note that the narrow range over which the $q^{-1}$ scaling is measured is intrinsic to the high flexibility/short persistence length of the polymer.

In the intermediate $q$ range, similarly to what is expected for semi-dilute linear polymer chains in a good solvent, a transition from a $q^{-5/3}$ decrease towards a pseudo-plateau at lower $q$ is measured. This transition gives a correlation length $\xi$ (usually referred as the blob size), which corresponds to the distance below which the polymer chain displays a single chain self-avoiding walk behavior [58]. This correlation length originates from time-dependent fluctuations of concentrations and can be satisfyingly determined using the scattering intensity of a semi-dilute polymer solution in a good solvent modeled by $I \sim \frac{(1+q\xi)^{1/3}}{(1+q^2\xi^2)}$ [59]. Note that this equation directly derives from a Lorentzian scaling law. Interestingly, our data show that the plateau regime does not persist at low $q$. Instead, the intensity is measured to vary as $q^{-2}$ in a broad range of wave-vectors that span two orders of magnitude. This low-$q$ behavior indicates the presence of large scale heterogeneities characterized by a fractal dimension of 2. This fractal dimension is consistent with monodisperse branched polymers in a good solvent [60]. Similar scattered intensity spectra (in the small and intermediate $q$-range) were predicted [61] and measured [59, 62-67] for swollen randomly cross-linked polymer networks and other polymeric gels. Biopolymer gels such as cellulose gels showed also similar profiles [68]. In these swollen systems, two kinds of heterogeneities are observed: the time-dependent concentration fluctuations which determine blob size, like in semi-dilute linear polymer solutions, and the large-scale static concentration heterogeneities due to the formation of "hard-to-swell zones" resulting from the heterogeneous crosslink density. The size of static heterogeneities, $\Xi$, could be eventually determined from the $q$-value from which the scattered intensity saturates at low $q$. However this size could not be measured in our experiment since the $q^{-2}$ dependence of the scattered intensity is found to hold till the smallest accessible $q$-vector, $q_{min}$=0.002 nm$^{-1}$. From our data, one can therefore only provide a lower estimate of the size of the heterogeneities, as $\Xi > 2\pi/q_{min} \approx 1000$ nm.

By analogy with the analysis performed in the literature [63], in the limit of $q\Xi \gg 1$, we propose to fit our experimental data with the following equation:

$I = a_1 \frac{(1+q\xi)^{1/3}}{(1+q^2\xi^2)} [1 + a_2 q^{-d_f}]$  (Eq. 1)

where $\xi$ is the correlation length of the semi-dilute polymer solution, $d_f$ is the fractal dimension measured at low $q$ and $a_1$ and $a_2$ are two scaling factors. Equation (1) accounts for the experimentally observed power law decrease at low $q$ and for the signature of a semi-dilute polymer solution in good solvent at intermediate and large $q$. It holds for length scale larger than the persistence length ($ql_0<1$) but smaller than the size of the heterogeneities $\Xi$ ($q\Xi>1$). Equation 1 can reproduce very well our experimental data measured by SAXS, as can be seen from the fits shown in Figure 10 for various sample concentrations ranging from 131 mg/ml to 393 mg/ml. The variation with the protein concentration of the two fitting parameters, $d_f$ and $\xi$, is displayed in Figure 11. We find that, within error bars, $d_f$ is constant over the whole range of concentrations: $d_f = 2.00 \pm 0.14$. This fractal dimension suggests one-phase network contrary to gels characterized by a q$^{-4}$ slope indicating the presence of sharp interfaces revealing arrested biphasic systems [69]. In the same range of concentration, $\xi$ is measured to continuously decrease with $C$, from 3 to 1 nm. A power law fit of the data gives $\xi \sim C^{-p}$, with $p$=0.8. Interestingly, the exponent is very close to the exponent predicted by scaling theories for semi-dilute polymer solutions in good solvent ($p$=3/4). The scaling theory [58] predicts $\xi = l_0 \phi^{-3/4}$, where $l_0$ is the polymer persistence length, or monomer size for flexible polymer, and $\phi$ is the volume fraction of polymer. In our case, we can evaluate the volume fraction of polymer from $C$, taking 1.31 for the dry density of protein [70]. A fit of the experimental data imposing $p = 3/4$ leads to $l_0 = 0.6$ nm, which is very similar to the value obtained above by studying the cross-over form a $q^{-1}$ to a $q^{-5/3}$ decay of the scattered intensity (inset Fig. 9). This totally validates the use of polymeric models. Moreover, this numerical value is in excellent agreement with the values experimentally found for unstructured proteins (between 0.5 and 0.7 nm) [71].

### 3.3 Discussion

The light scattering data in the very dilute regime shows that the scattering objects are close to monodisperse, and have an apparent molecular weight far larger than the largest one present in the protein mixture (Fig. 1). They presumably contain more than one protein molecule, and are instead complexes comprising several gluten proteins. One can calculate the concentration of proteins in the scattering objects, as $\frac{3M_w^{app}}{4\pi \mathcal{N}_a R_G^3}$, where $\mathcal{N}_a$ is the Avogadro number, $M_w^{app}$ is the apparent molecular weight, and $R_G$ the radius of gyration. We find a concentration of the order of 3.2 mg/ml, showing that the complexes are very little dense. In accordance, they exhibit internal dynamics once they are probed at length scales smaller than their radius of gyration. Interestingly, the internal dynamics exhibits all the hallmarks theoretically expected and experimentally found for polymer chains in good solvent. In addition, our data suggest that the complexes are not linear chains but instead cross-linked or branched polymer chains. Accordingly, we find that the radius of gyration of the objects is smaller than the one measured for linear flexible chains of equivalent molecular weight [72]. However, light scattering measurements do not allow one to establish whether the protein assemblies that we measure comprise both glutenin and gliadin proteins, or uniquely glutenin proteins. If gliadin and glutenin proteins do not interact and scatter independently, one could evaluate the respective amplitude of their scattering. The intensity is proportional to $NR^6$, where $N$ is the number of scattering objects, and $R$ is the radius of the scattering objects. Our protein extract comprises about 50% w/w gliadin and 50% w/w glutenin. Assuming that the protein assemblies measured (molecular weight 27000000 g/mol, hydrodynamic radius of the order of 130 nm) are uniquely made of glutenin and that gliadin (hydrodynamic radius of the order of 10 nm -data not shown- and average molecular weight of 40000 g/mol) scatter independently, one predicts that, at low $q$, the scattering from the small gliadins would be about 1000 times smaller than the one of the big objects. Hence, this scattering could not be detected. Nevertheless, based on calorimetric analysis of gluten and its constituting proteins, gliadin and glutenin can be viewed as perfectly blended at the molecular level [73]. Moreover, rheological investigations of gluten and gliadin blends supports the view that gliadin interferes with glutenin-glutenin interactions [74]. Hence, we cannot exclude that the protein complexes we probe in ethanol/water comprise gliadin in addition to glutenin. In addition, our data suggest also that the objects are rather monodisperse, which is quite intriguing. A tentative explanation might be found based on an analogy with colloidal systems, where monodisperse clusters have been observed and predicted to occur when the interaction potential between the colloidal particles exhibit a short-range attractive potential and a long-range repulsive potential due to electrostatic [75, 76]. This point however certainly needs further investigation.

Wide-angle X-ray scattering (WAXS), small-angle X-ray scattering (SAXS) and very small-angle neutron scattering (VSANS) experiments, which probe the structure of the gluten protein complexes in the concentrated regime, at length scales ranging from 0.3 to 1000 nm, provide results in full agreement with the findings in the dilute regime. In particular, at small length scale, but still larger than the persistence length, the scattered intensity is found to varies as a power law of the wave-vector with an exponent -5/3, which is precisely the exponent expected for a self-avoiding polymer chain in a good solvent. Note that the scaling has been recently measured for another intrinsically disordered protein and interpreted similarly to us [77]. A scaling close to $q^{-5/3}$ has also been measured for zein, a storage protein from maize [78] although an alternative interpretation has been given. In our experiments, the $q^{-5/3}$ scaling is found to hold up to a certain length scale, $\xi$. We measure in addition that this correlation length scale decreases when the protein concentration increases as $C^p$ with $p$=0.8, a value very close to 3/4, the theoretically expected value for polymer in good solvent in the semi-dilute regime. For length scale larger than $\xi$, the scattered intensity is found to increase strongly as the wave-vector decreases, indicating the presence of large length scale heterogeneities in the samples, as observed for polymeric gels [66, 79], polysaccharide gels [68, 80], polymeric coacervates [81] and also denaturated globular proteins gels [69, 82]. In our case, the concentrated samples display rich and complex visco-elastic properties which are directly related to this hierarchical organization of the protein chains [83].

Altogether, our experiments provide several quantitative evidence that the gluten protein mixture behaves in the ethanolic solvent used here as rather standard flexible polymeric chains in good solvent. This finding is at odds with results obtained using multiangle laser light scattering coupled to size exclusion chromatography (SEC-MALLS). In SEC-MALLS, the scaling between the radius of gyration and the molecular weight $R_G \sim M_w^{\upsilon}$ is measured. Several previous studies [84-86] using glutenin proteins in various solvents (including acetic acid, phosphate and citric buffers with eventually SDS), suggest an exponent $\upsilon$ of 1/3 or even lower. An exponent of 1/3 is expected for compact objects, whereas an exponent lower than 1/3 is unphysical. By contrast our data indicate a much more open structure ($\upsilon$ =0.6 for polymer coils in good solvent conditions). The discrepancy between those measurements and our measurements could originate from some specific interactions between the proteins (or some of the proteins of our extract) and the chromatography columns used for the SEC-MAALS measurements, or from the fact that the scattering objects are branched (as shown by us). Interestingly, our results can instead be related with structural investigations by scattering techniques of dilute aqueous suspensions (comprising 20 mM acetic acid) of the hydrophilic central repetitive domain of a high molecular weight subunit of gluten protein [87]. The authors demonstrate that the scattering data are in agreement with a structural model of flexible cylinder in a good solvent at odds with previous claims of stiff objects [88, 89]. Similarly to us, they found that, for length scale between the persistence length and the radius of gyration, the scattering intensity scales as $q^{-5/3}$. By measuring the high-$q$ cut-off of the $q^{-5/3}$ regime, they can estimate a persistence length of the order of 1.5 nm. This value is in rough agreement with our value, (0.6-0.7) nm, for complex gluten protein mixtures in ethanol/water solvent, although our data suggest a slightly more flexible structure.

Hence, although glutenin proteins are very often modeled as a linear polymer back-bone of glutenin sub-units linked together by disulfide bonds [3, 4], we believe our experimental work provide the first direct experimental evidence that the dynamics and structure of those proteins can be qualitatively and quantitatively accounted by simple models developed for synthetic polymers in good solvent in water/ethanol mixtures.

## 4. Conclusion

We have investigated a complex extract of wheat gluten proteins, comprising both monomeric and polymeric proteins, and exhibiting broad size distribution of molecular weight and structures. The protein extract can be dissolved in a water/ethanol solvent for a wide range of protein concentrations. We have found that stiff gels form for large protein concentrations. By studying the structure and dynamics both in the dilute regime and in concentrated regime we have shown that the sample behave as flexible polymers in a good solvent. We have also some experimental evidence that those polymers are branched. By combining several techniques, we have probed the structure of the polymeric gel in a large range of length scale form nm to μm and evidenced that the proteins self-assemble in a hierarchical fashion. This system appears therefore as a unique model system to investigate the supramolecular organization of gluten proteins and its impact on the viscoelastic properties of gluten gels. We are currently investigating the rheological properties of those gels. Other important open questions include the respective role of the two solvents (water and ethanol) and the nature of the interactions between gliadins and glutenins.

## Acknowledgements

We thank Adeline Boire (IATE) for advice in the protein extraction, Joelle Bonicel (IATE) for help in the SE-HPLC, Philippe Dieudonné (L2C) for the in-house X-Ray scattering measurements and Céline Charbonneau (L2C) for help during the VSANS measurements. We acknowledge the European

Synchrotron Radiation Facility (ESRF), Grenoble, France for provision of synchrotron radiation facilities. We would like to thank Pawel Kwasniewski for assistance in using beamline ID02 at ESRF and Zhendong Fu for assistance during the VSANS measurements at the Heinz Maier-Leibnitz Zentrum (MLZ), Garching, Germany. Financial supports from Laboratoire of Excellence NUMEV (ANR-10-LAB-20) and from INRA (CEPIA department) are acknowledged. This research project has been supported by the European Commission under the 7th Framework Program through the "Research Infrastructures" action of the Capacities Program, NMI3-II, Grant Agreement number 28388 to perform the neutron scattering measurements at MLZ.

## Supporting information

We provide SE-HPLC profiles of the different gluten extracts, and of the native gluten and the fraction of interest (C2). This information is available free of charge via the Internet at http://pubs.acs.org

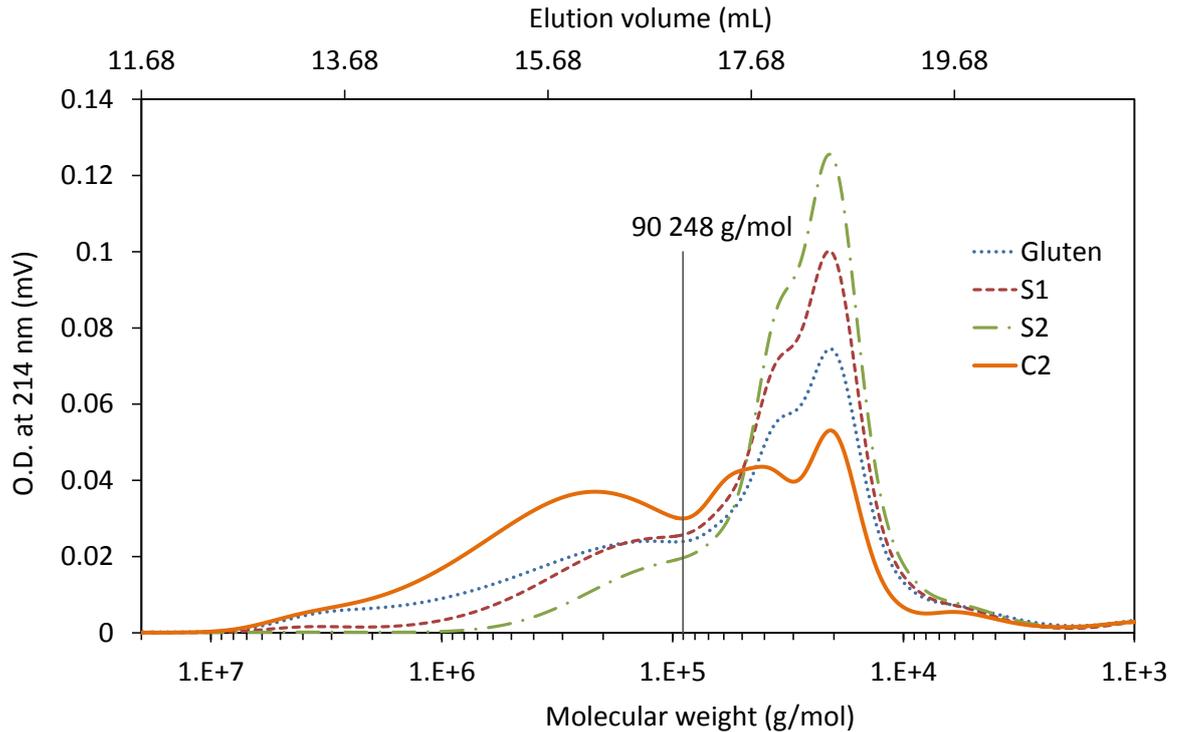

**Figure 1**: SE-HPLC profiles of the different gluten extracts. The profiles are for equivalent total protein content (9.8 µg). S1 : first water/ethanol soluble extract from gluten; S2 : light- and C2 : dense-phase obtained after liquid/liquid phase separation of S1 at 4°C. Chromatography was performed on two columns placed in series (TSK gel G6000 and G5000 PWXL, each 30 cm x 7.8 mm I.D., TosoBioscience) and preceded by a guard-column (TSK gel PWXL, 4 cm x 6 mm I.D., TosoBiocience). Relative molecular weight calibration was obtained from known protein standards (see Material & Methods section).

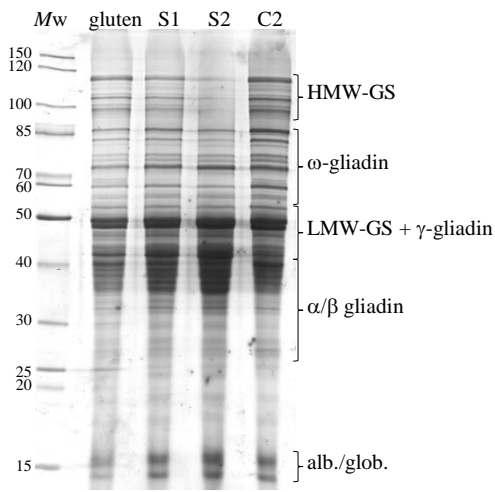

**Figure 2**: SDS-PAGE of the different gluten extracts. S1: first water/ethanol soluble extract from gluten; S2: light phase and C2: dense phase obtained after liquid/liquid phase separation of S1 at 4°C. $M_W$: molecular weight standards in kDa (PageRuler$^{TM}$ unstained protein lader, Thermo Fisher). SDS-PAGE (13%T-0.63%C) is performed in reduced conditions. Each lane was loaded with 30µg of protein.

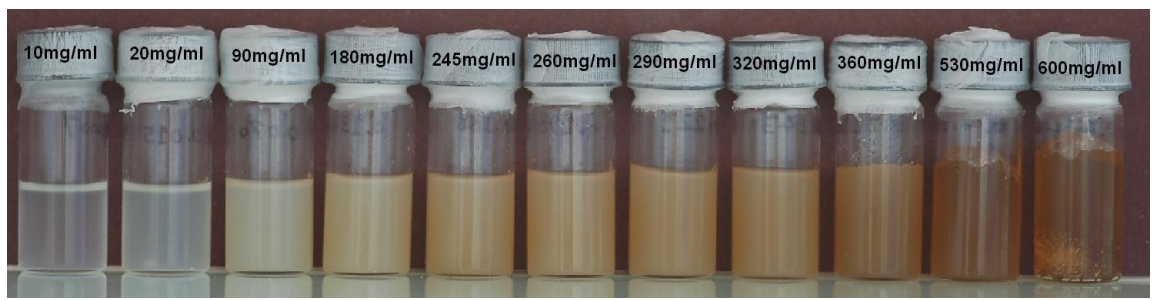

**Figure 3**: Picture of samples with protein concentrations, *C,* from 10 mg/ml to 600 mg/ml, in a 50/50 v/v water/ethanol mixture.

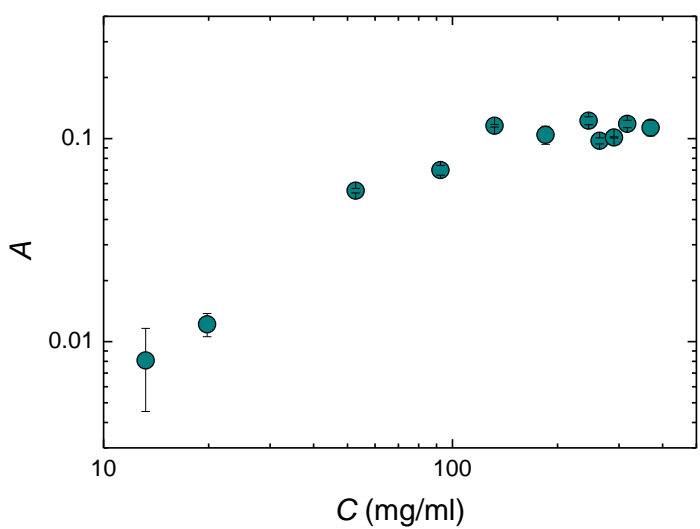

**Figure 4**: Sample absorbance as measured at wavelength 600 nm as a function of the protein concentration.

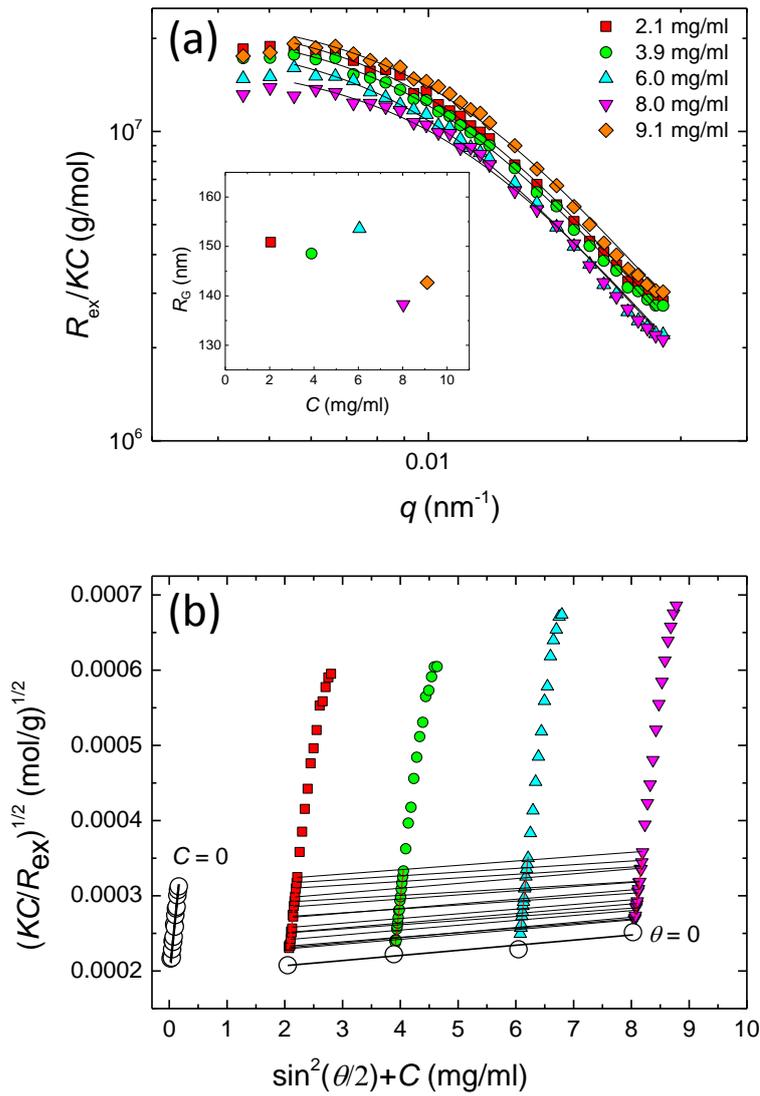

**Figure 5:** Static light scattering measurements for diluted samples with various concentrations of proteins as indicated in the legend. (a) Scattered intensity as a function of the wave-vector (Inset) Radius of gyration, as evaluated from a Debye fit, as a function of the protein concentration. (b) Berry plot of the same data as in (a). Symbols are experimental data and lines are best fits using a Debye function (a) and linear fits (b).

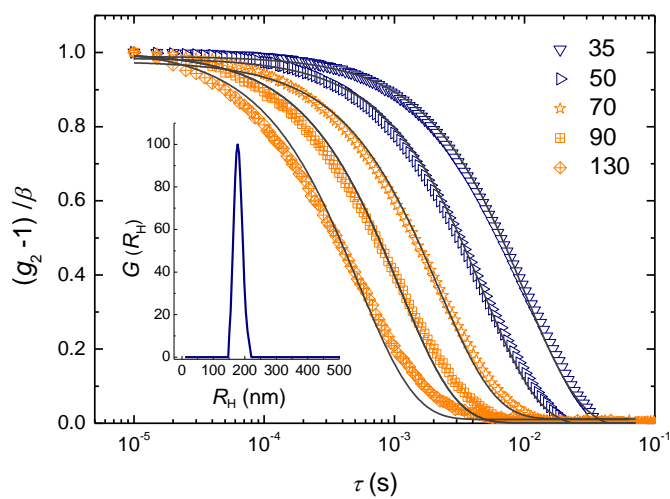

**Figure 6:** Correlation functions measured at different angles, $\theta$, as indicated in the legend, for a sample with protein concentration $C = 6.0$ mg/ml. The symbols are the experimental data points and the lines are best fits with a single exponential decay functional form. (Inset) Size distribution measured at $\theta = 35$ deg as deduced from a Contin analysis.

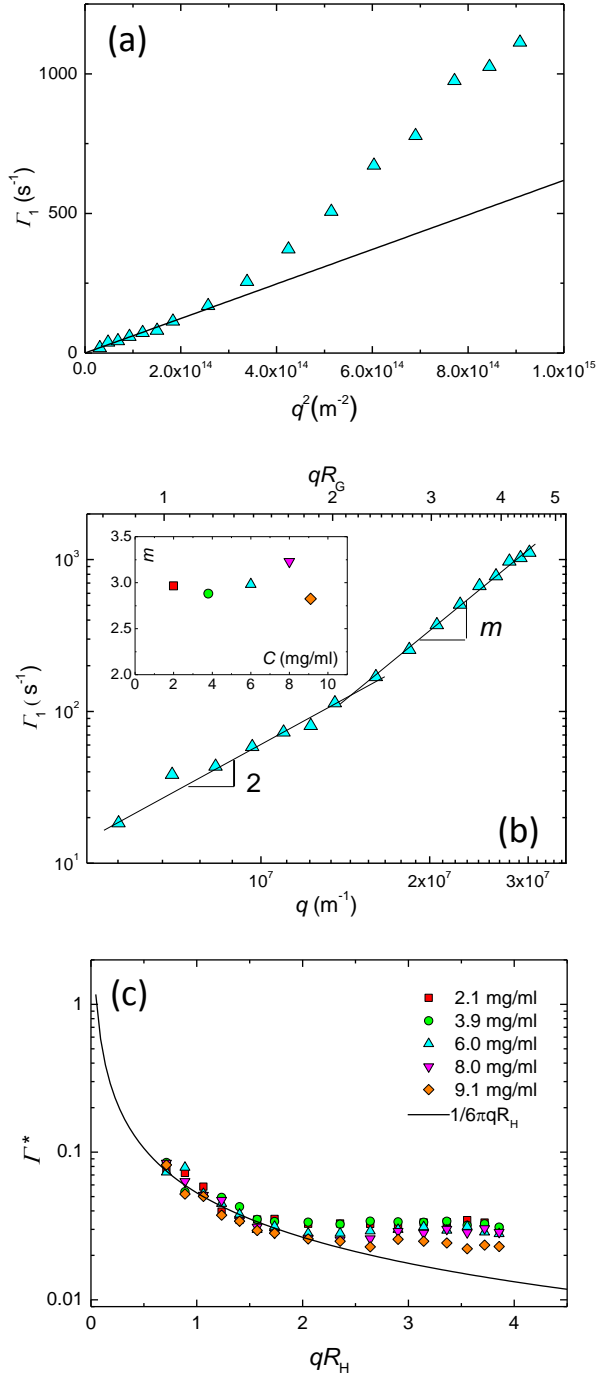

**Figure 7**: (a,b) Decay rate, $\Gamma_1$, as measured from a cumulant analysis of the correlation functions for a sample with protein concentration $C$=6.0 mg/ml. In (a), $\Gamma_1$ is plotted as a function of $q^2$; symbols are experimental data points and the line is a linear fit of the low-$q$ data points. In (b), $\Gamma_1$ is plotted as a function of $q$ (bottom axis) and $qR_G$ (top axis); symbols are experimental data points and the lines are power law fit with exponent 2 at low $q$ and with exponent $m$ at high $q$. (Inset) Plot of $m$ for the different protein concentrations. (c) Plot of $\Gamma^* = \frac{\Gamma_1}{q^3}\frac{\eta_0}{k_BT}$, where $k_BT$ is thermal energy and $\eta_0$ is the solvent viscosity, as a function of $qR_H$, where $R_H$ is the hydrodynamic radius. Symbols are the values computed from the decay rate for different protein concentrations as indicated in the legend and the line is the theoretical expectation for a purely translational diffusion.

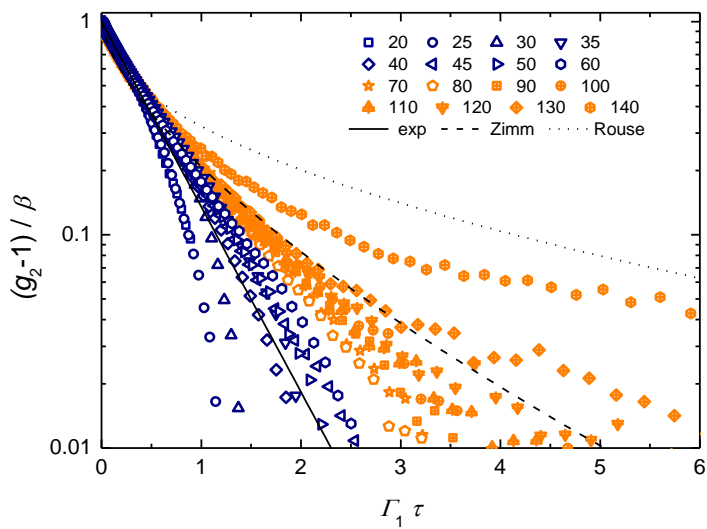

**Figure 8:** Correlation functions plotted as a function of $\Gamma_1\tau$, where $\Gamma_1$ is the decay rate extracted from a cumulant analysis, for different scattering angles, as indicated in the legend. The protein concentration is $C = 6.0$ mg/ml. The correlation functions measured at low angles ($qR_G<2$) are plotted in dark blue and those measured at larger angles ($qR_G>2$) are plotted in orange. Lines are theoretical expectations for different models, as explained in the text.

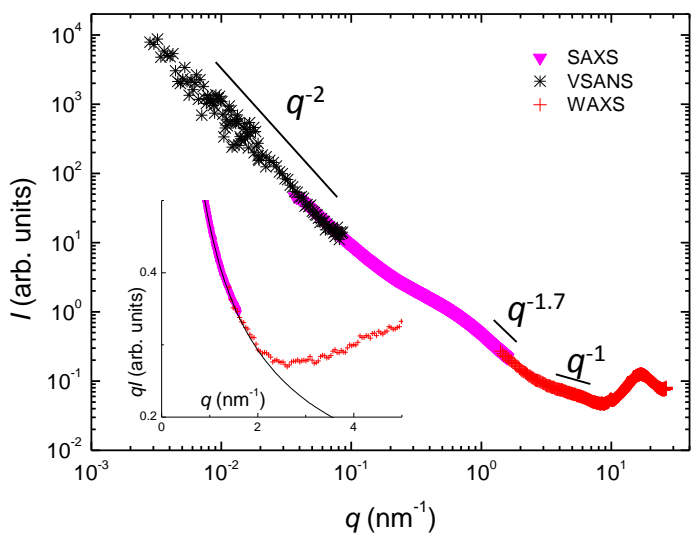

**Figure 9:** Scattered intensity as a function of the wave-vector as measured by wide-angle X-ray scattering (red crosses), small-angle X-ray scattering (pink triangles), and very-small neutron scattering (black stars), for a sample with a protein concentration $C=290$ mg/ml. (Inset) Holtzer plot, $qI$ vs $q$, of the same data as in the main plot. Only data in the high $q$ range are shown. The black line is a $q^{-5/3}$ power law.

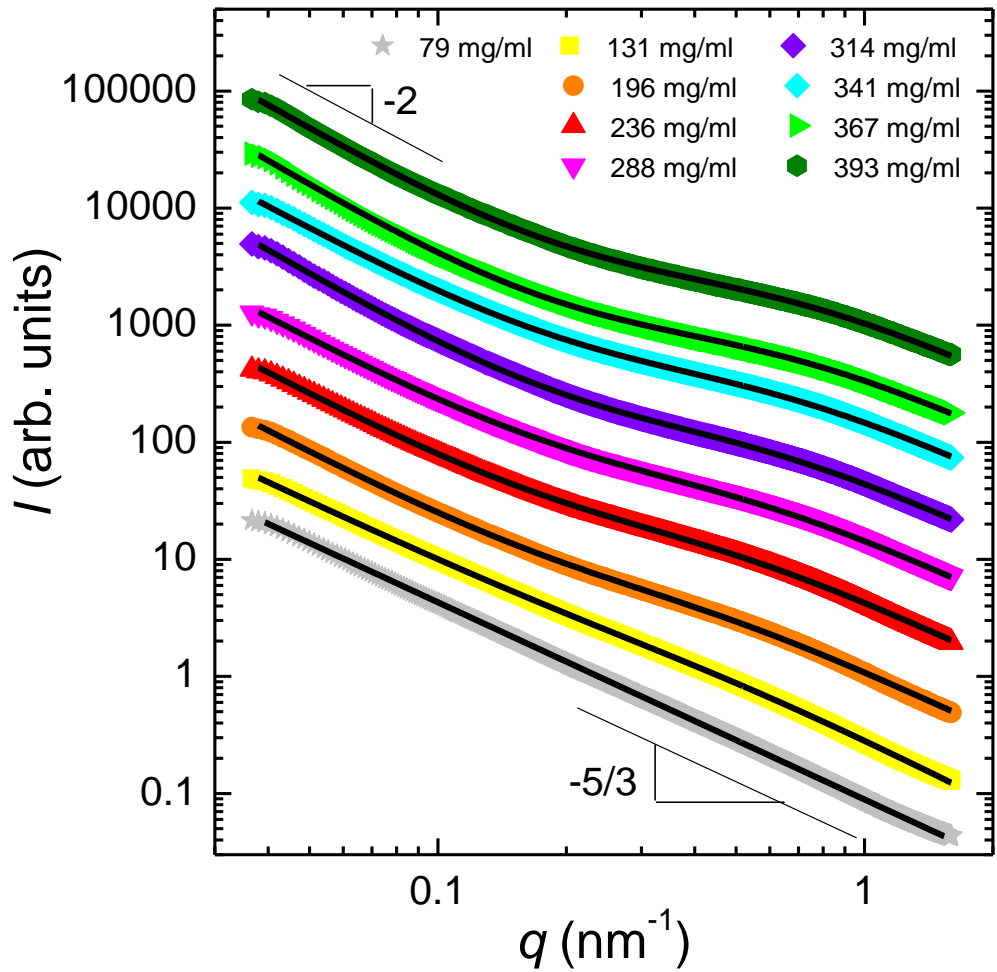

**Figure 10:** Scattered intensity, *I*, as a function of the wave-vector, *q*, as measured by small-angle X-ray scattering, for various protein concentrations as indicated in the legend. The data have been arbitrarily shifted vertically. Symbols are experimental data points. The black lines are best fits, using the model described in the text (Eq. 1) for concentrations *C* in the range (131-393) mg/ml, and using a power law with an exponent -5/3 for *C* = 79 mg/ml (gray symbols).

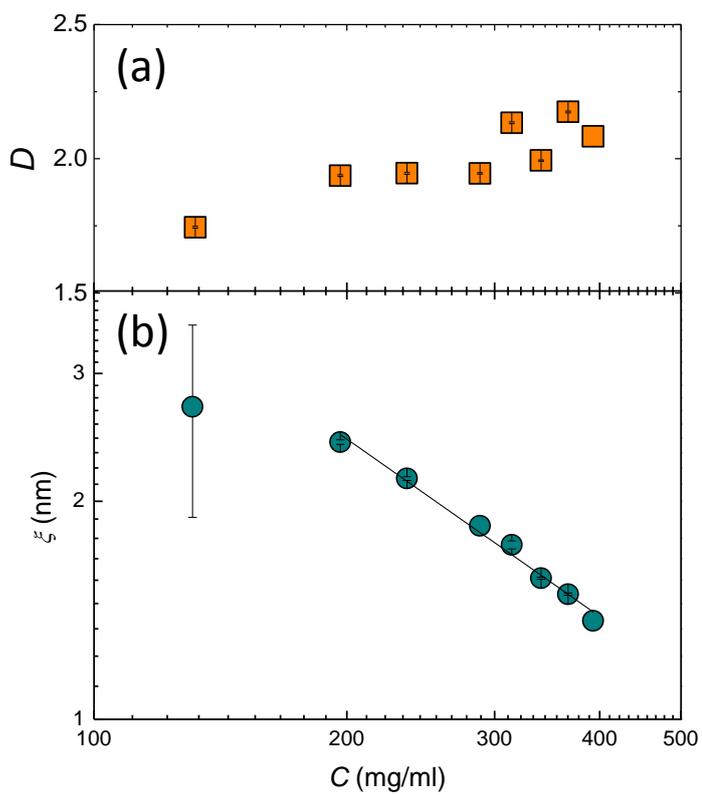

**Figure 11:** (a) Fractal dimension and (b) Correlation length, as determined from the fit of the SAXS profile as a function of the protein concentration (Fig. 10). Symbols are experimental data points and the line in (b) is a power law fit yielding an exponent of -0.8.

**Supplementary materials**

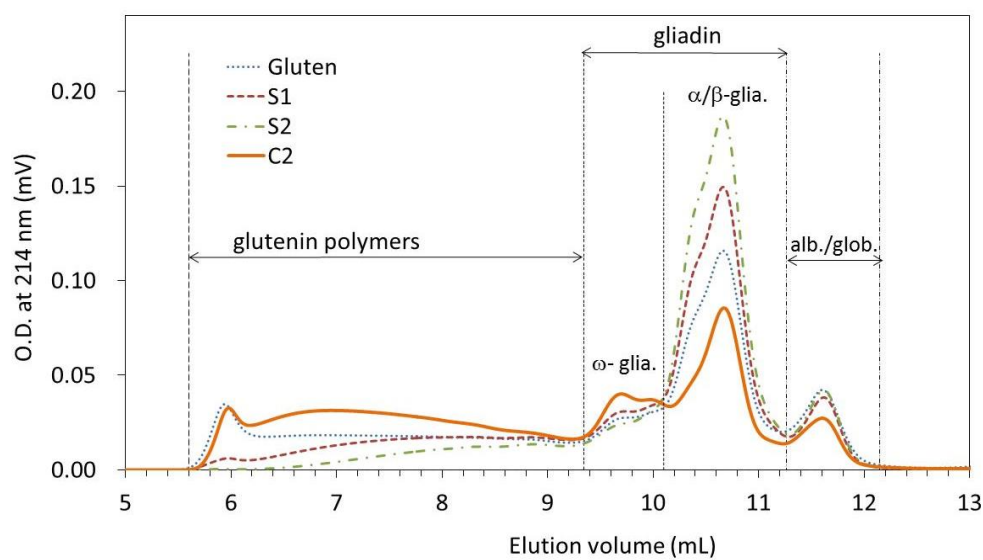

**Figure S1:** SE-HPLC profiles of the different gluten extracts. The profiles are for equivalent total protein : first water/ethanol soluble extract from gluten; S2 : light phase and C2 : dense phase obtained after liquid/liquid phase separation of S1 at 4°C. Chromatography was performed on a TSK G4000 SWXL column (30 cm x 7.8 mm I.D., TosoBioscience). Relative molecular weight calibration was obtained from known protein standards (as detailed in Material & Methods section).

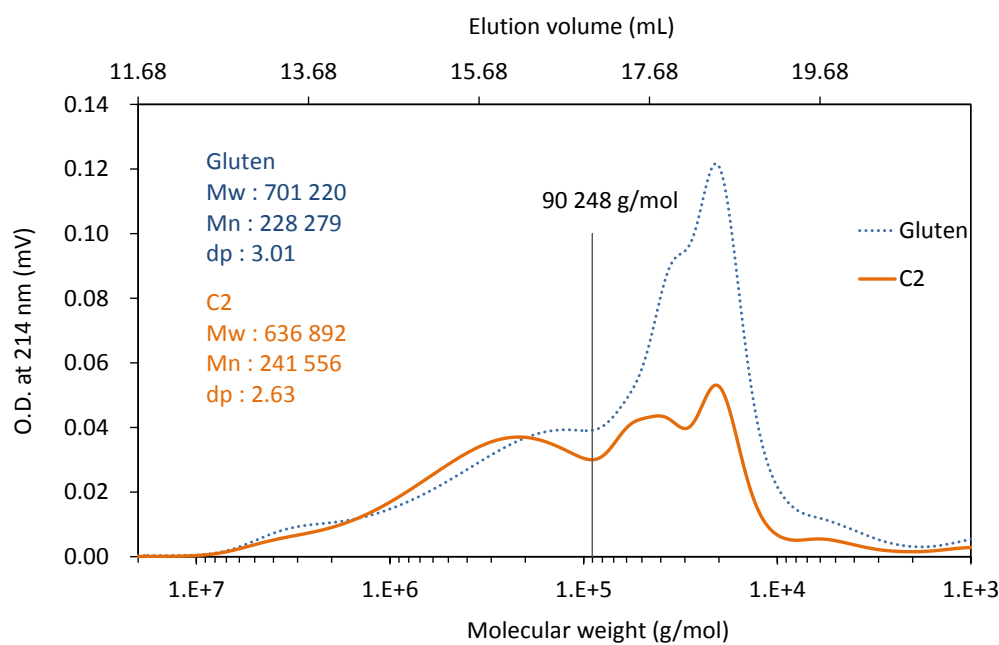

**Figure S2:** SE-HPLC profiles of the native gluten and the fraction of interest (C2). The profiles have been normalized to have the same total amount of glutenin. Chromatography was performed on two columns placed in series (TSK gel G6000 and G5000 PWXL, each 30 cm x 7.8 mm I.D., TosoBioscience) and preceded by a guard-column (TSK gel PWXL, 4 cm x 6 mm I.D., TosoBiocience). Relative molecular weight calibration was obtained from known protein standards (see Material & Methods section).